\documentclass[
   ,final            
  ]
  {aipproc}

\layoutstyle{6x9}

\begin{document}

\title{Short-Bright GRBs: spectral properties}

\author{G. Ghirlanda}{
  address={IASF - Via Bassini 15, I-20133 Milano, Italy.}
}
\author{G. Ghisellini}{
  address={OAB - Via Bianchi 46, I-23807 Merate, Italy.}
}
\author{A. Celotti}{
  address={SISSA - Via Beirut 2, I-34014 Trieste, Italy.}
}

\begin{abstract}
We study the spectra of short--bright GRBs detected by BATSE and
compare them with the average and time resolved spectral properties of
long--bright bursts. We confirm that short events are harder than long
bursts, as already found from the comparison of their (fluence) hardness
ratio, but we find that this difference is mainly due to a harder low
energy spectral component present in short bursts, rather than to a
(marginally) different peak energy. Moreover, we find that short GRBs
are similar to the first 1 sec emission of long bursts.  The
comparison of the energetic of short and long bursts also suggests
that short GRBs do not obey the peak energy--equivalent isotropic
energy correlation recently proposed for long events, implying that
short GRBs emit lower energy than long ones. Nonetheless, short bursts
seem to emit a luminosity similar to long GRBs and under such
hypothesis their redshift distribution appears consistent with that
observed for long events. These findings might suggest the presence of
a common mechanism at the beginning of short and long bursts which
operates on different timescales in the these two classes.
\end{abstract}

\maketitle

\section{Introduction}

Evidences supporting the possible different nature of short and long
GRBs are: (1) their bimodal duration distribution
(\cite{Kouveliotou1993}) with mean duration of $\sim 20$ sec and $\sim
0.3$ sec for long and short events, respectively; (2) their different
temporal properties (e.g. number and width of pulses in the light
curve, \cite{Norris2000}, \cite{Nakar2002}, \cite{McBreen2001}); (3)
the larger hardness ratio of short bursts (\cite{Kouveliotou1993},
\cite{Tavani1998}, \cite{Paciesas2001}). Moreover, theoretical models
for the progenitors of GRBs associate short bursts to the merger of
compact objects in a binary system (\cite{Goodman1986},
\cite{Meszaros1992}) while long GRBs seem to be connected to the
core--collapse of massive stars (\cite{Woosley1993},
\cite{Vietri1998}).

Despite the increasing understanding of the nature of long duration
$\gamma$--ray bursts, the population ($\sim 1/3$) of sub-second short
GRBs is still largely not understood.  This is mainly due to (i) the
low signal-to-noise ratio which strongly limits the analysis of the
prompt (temporal and spectral) emission of short events, and (ii) the
lack of any firm afterglow measurement for short GRBs, which indeed
represented a major advance in unveiling the mystery of long bursts.

\section{Short vs. Long: spectral properties}

The emission properties of long GRBs have been studied in details
(\cite{Band1993}, \cite{Preece2000}) by fitting their time average
(\cite{Band1993}) and time resolved (\cite{Preece2000},
\cite{Ghirlanda2002}) broad-band high-resolution spectra. Nonetheless,
the comparison with the spectral properties of short bursts has been
based mainly on the analysis of their fluence hardness ratio
(e.g. \cite{Cline1999}, \cite{Yi-ping Qin2001}) which is marginally
representative of the effective spectral shape. For this reason we
analyzed (\cite{Ghirlanda2003a}) a sample of bright--short BATSE
bursts (selected with peak flux $\ge 10$ phot/cm$^2$ sec between
50-300 keV) by fitting their $\gamma$--ray spectra ($\sim 30 - 1800$
keV) with the standard spectral functions
(e.g. \cite{Preece2000}). The set of spectral parameters (namely the
low energy photon spectral index $\alpha$ and the EF$_{E}$ peak energy
$E_{peak}$) were compared with those of long bright bursts
(\cite{Preece2000}) both considering time integrated and time resolved
spectra.

\begin{figure}
  \includegraphics[height=0.25\textheight,width=0.5\textwidth]{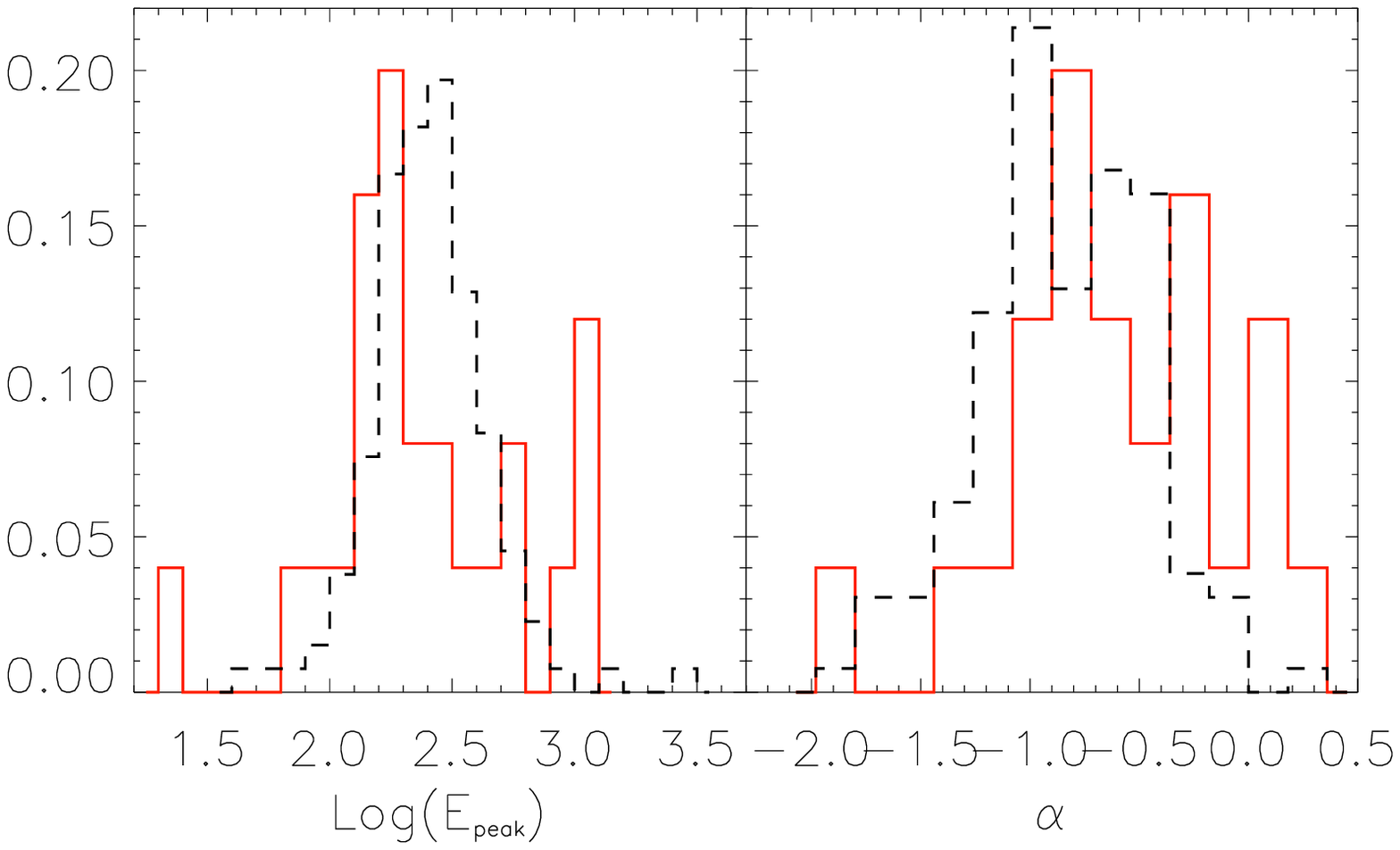}
  \includegraphics[height=0.25\textheight,width=0.5\textwidth]{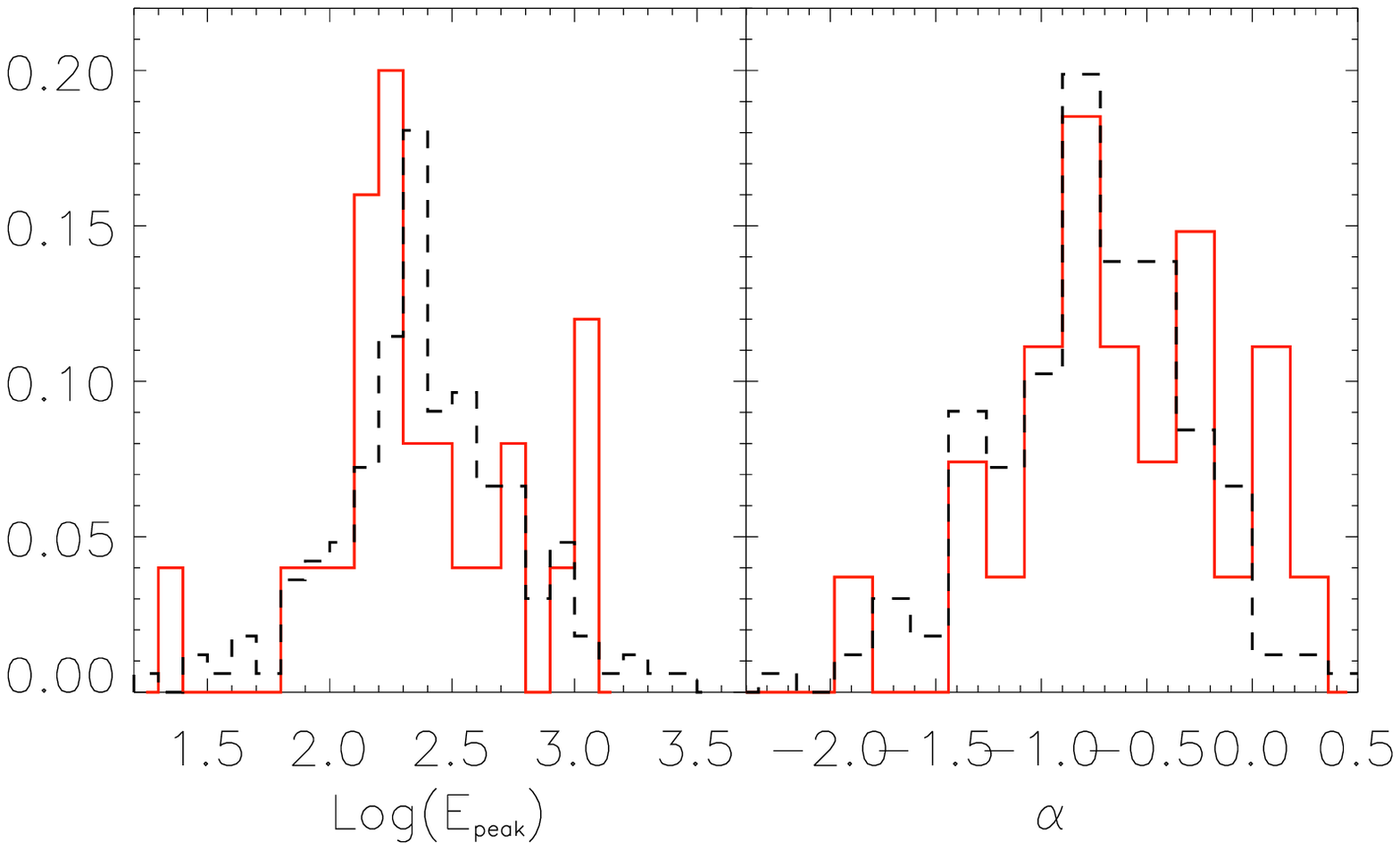}
  \caption{Distribution of the spectral parameters for short ({\it
  solid line}) and long ({\it dotted line}) GRBs. {\it Left}: peak
  energy ($E_{peak}$) of the EF$_{E}$ spectrum and low energy photon
  spectral index ($\alpha$) for the average spectra of long and short
  bursts. {\it Right}: $E_{peak}$ and $\alpha$ of the first 1 second
  of long bursts compared to the average spectra of short
  events.}\label{fig1}
\end{figure}

As shown in fig.\ref{fig1} ({\it Left}) short and long GRBs present
different $\alpha$ distributions and only marginally different
$E_{peak}$ distributions (with a small Kolmogorov-Smirnov test
probability - i.e. 4\% and 10\%, respectively - for the two population
to be similar). The average values are $\langle \alpha
\rangle=-0.84\pm 0.15$, $\langle E_{peak}\rangle=305\pm 22$ keV and
$\langle \alpha \rangle=-0.58\pm 0.10$, $\langle
E_{peak}\rangle=355\pm 30$ keV for long and short GRBs,
respectively. This suggests that short bursts have (average) harder
spectra than long events, as also indicated by the analysis of their
hardness ratio, \emph{but}, intriguingly, this difference is due to a
considerably \emph{harder low energy spectral component} present in
short bursts rather than to a different peak energy.
The comparison of short GRBs with the time resolved spectral
properties of long bursts shows that short events have spectra similar
to the first 1 sec of long GRBs, as shown in fig.\ref{fig1} ({\it
Right}), with a high KS probability of 23\% and 80\% for $\alpha$ and
$E_{peak}$ to be similar. This might suggest that a similar mechanism
operates for the complete duration of short bursts and in the first 1
sec of long GRBs.

\section{Short vs. Long: energy and luminosity}

The analysis of the intrinsic properties of (still few) long bursts
with known redshift (\cite{Amati2002}) highlighted a possible
correlation ($E_{iso}\propto E_{peak}^{1.93}$) between is the
equivalent isotropic burst energy $E_{iso}$ and the spectral peak
energy $E_{peak}$.  This correlation has been also confirmed by
Hete--II (\cite{Lamb2003}) and a similar relation has been found
between $E_{peak}$ and the burst isotropic luminosity $L_{iso}$
(\cite{Yonetoku2003}).
\begin{figure}
  \includegraphics[height=0.4\textheight,width=0.5\textwidth]{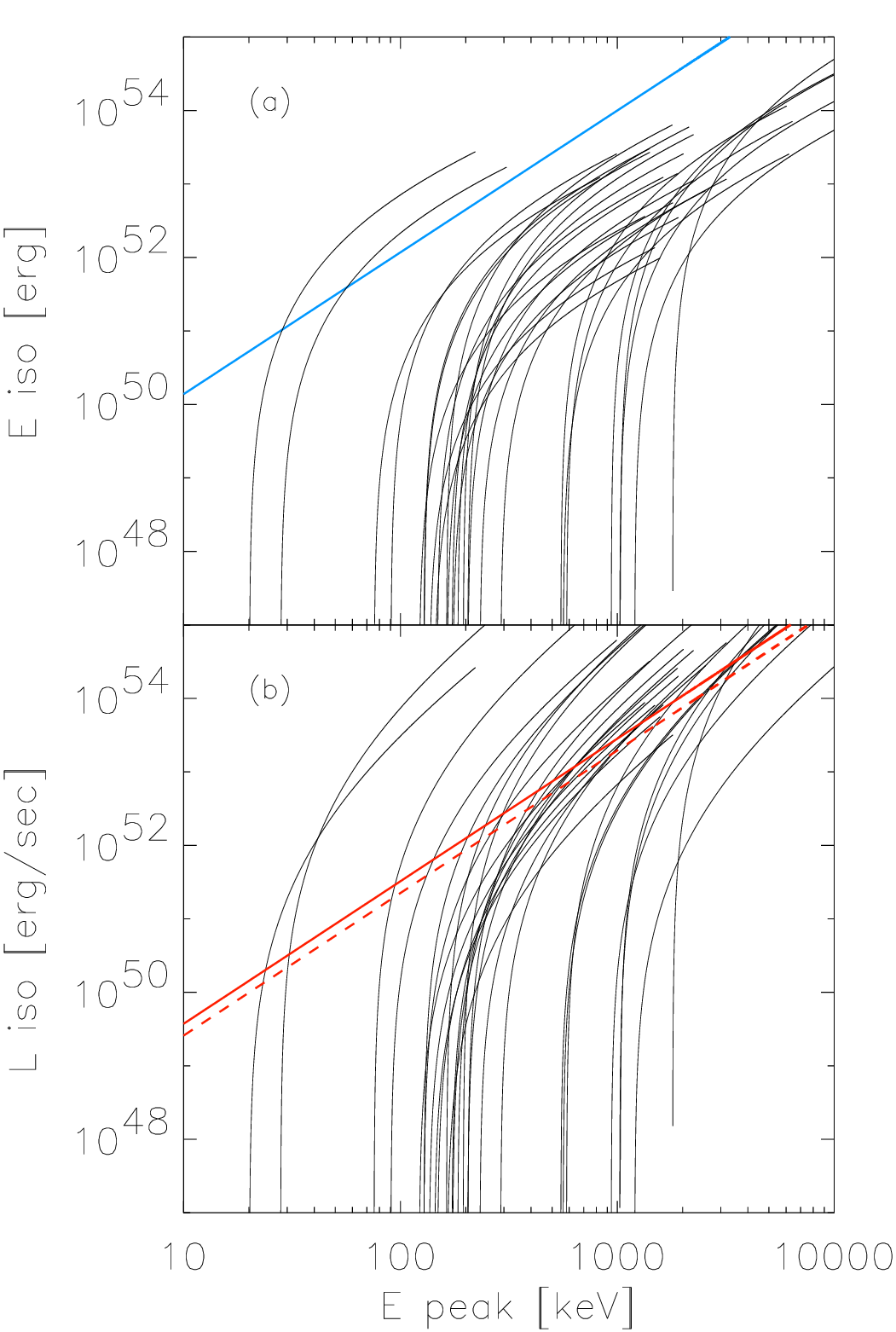}
  \includegraphics[height=0.4\textheight,width=0.5\textwidth]{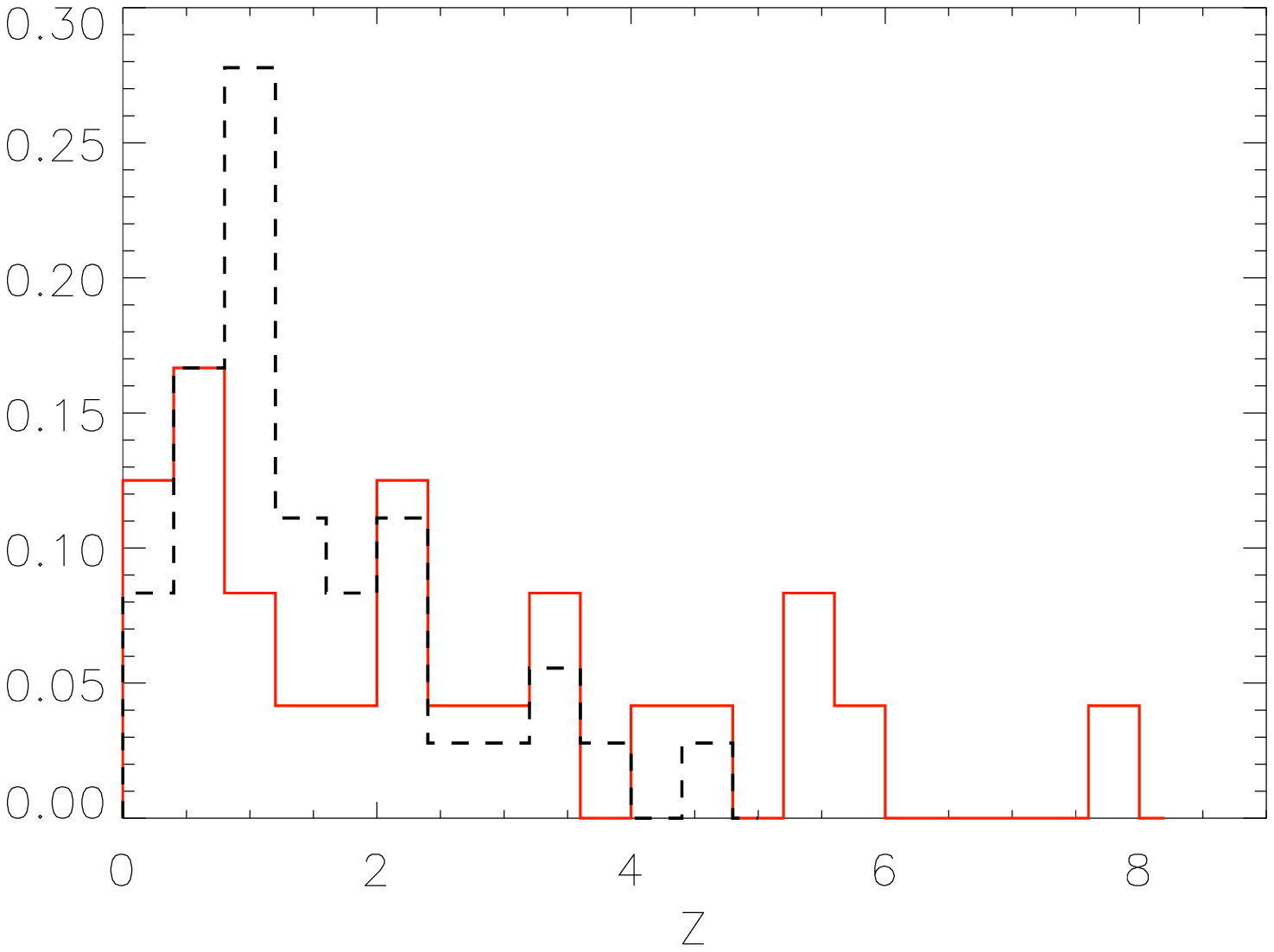}
  \caption{{\it Left}: Equivalent isotropic energy $E_{iso}$ ({\it a})
  and equivalent isotropic luminosity $L_{iso}$ ({\it b})
  vs. $E_{peak}$. Solid lines represent the correlations between these
  intrinsic properties found for long bursts (\cite{Amati2002} and
  \cite{Yonetoku2003}). Curves represent the intrinsic $E_{iso}$
  ($L_{iso}$) and $E_{peak}$ for short bursts assuming a variable
  redshift between 0.1 and 10. {\it Right}: Redshift distribution of
  short bursts ({\it solid line}) assuming that they indeed satisfy
  the $L_{iso}$ vs. $E_{peak}$ relation. For comparison also the
  observed distribution of long bursts ({\it dotted line}) is
  reported.}\label{fig2}
\end{figure}
If short bursts were similar to long events they might satisfy these
correlations between $E_{iso}$ ($L_{iso}$) and $E_{peak}$. However, no
redshift of short GRBs has been measured so far. Nonetheless, we can
still verify (\cite{Ghirlanda2003a}) the above hypothesis assuming a
variable redshift (between 0.1 and 10) and scaling the observed
spectral properties of the sample of bright short bursts in the source
rest frame.

Fig.\ref{fig2}{\it-a} shows that for any assumed $z$ short bursts
(solid curves) populate a region below the $E_{peak}$ - $E_{iso}$
correlation of long bursts (\cite{Amati2002} - solid line). On the
other hand (fig.\ref{fig2}{\it-b}) the luminosity of short events is
consistent with the proposed relation for long bursts
(\cite{Yonetoku2003} - solid line). In conclusion short and long
bursts seem to emit a \emph{similar equivalent isotropic luminosity}
but different (lower in short bursts) energy due to their different
duration. Nonetheless, short and long bursts might still have similar
emitted energies if short bursts are less collimated than long events
although, in this case, short bursts would have a much larger
luminosity.  Under the first hypothesis we can further extract from
the $L_{iso}$-$E_{peak}$ relation the possible redshift distribution
of short bursts (fig.\ref{fig2}-{\it right}, {\it solid line}). This is
consistent with the observed $z$ distribution of long events
(fig.\ref{fig2}-{\it right}, {\it dotted line}) and again supports a
possible similarity of short and long GRBs.

\section{Discussion}

The comparison of the spectral properties of short and long bursts
pointed out that short bursts are harder than long events due to a
(average) harder low energy spectral component (rather than to a
different peak energy). The spectra of short bursts are similar to the
first 1 sec of the emission of long bursts. Intriguingly, short bursts
present a similar intrinsic luminosity but a lower (isotropic) energy
than long bursts if the recently found correlations between these
quantities were true also for short GRBs. Under such hypothesis the
implied redshift distribution of short bursts results similar to that
observed in long GRBs, and this prediction will be tested in the
forthcoming Swift era. Nonetheless, short GRBs might still have a
similar energy to that of long bursts if their collimation angle is
much larger than that of long events, but in this case their intrinsic
luminosity should be much larger than that of long bursts.  These
results suggest the presence of a common mechanism operating at the
beginning of short and long bursts which could explain their similar
spectral properties and luminosity.  If this is the case a possible
difference in the burst dynamical evolution (e.g. fallback of the
pre-GRB ejected material) might play a crucial role in distinguishing
these two classes.

\begin{theacknowledgments}
G. Ghirlanda would like to thank the organizing committee for the full
grant support for the participation to this conference.
\end{theacknowledgments}

\bibliographystyle{aipprocl} 

\begin{thebibliography}{}

\bibitem{Amati2002} Amati L. et al., 2002, A\&A, 390, 81
\bibitem{Band1993} Band D.L., et al., 1993, ApJ, 413, 281
\bibitem{Cline1999} Cline D. B., Matthey C., Otwinowski S., 1999, ApJ, 527, 827
\bibitem{Ghirlanda2002} Ghirlanda G., Celotti A. \& Ghisellini
G., 2002, A\&A, 393, 409
\bibitem{Ghirlanda2003} Ghirlanda G., Celotti A. \& Ghisellini
G., 2003, A\&A, 406, 879
\bibitem{Ghirlanda2003a} Ghirlanda G., Ghisellini G., Celotti A.,
astro-ph/0310861 
\bibitem{Goodman1986} Goodman J, 1986, ApJ, 308, L47
2003, astro-ph/0211620
\bibitem{Kouveliotou1993} Kouveliotou K. et al., 1993, ApJ, 413, L101
\bibitem{Lamb2003} Lamb D., 2003a, astro-ph/0309462
\bibitem{McBreen2001} McBreen S. et al., 2001, A\&A, 380, L31
\bibitem{Meszaros1992} M\'esz\'aros P. \& Rees M. J., 1992, ApJ,
397, 570
\bibitem{Nakar2002} Nakar E. \& Piran T., 2002, MNRAS, 330, 920
\bibitem{Norris2000} Norris J., Scargle J. \& Bonnel J., 2000,
BAAS,32,1244
\bibitem{Paciesas2001} Paciesas et al., 2001, Woods Hole conf. proc.
Ricker \& Vanderspek
\bibitem{Preece2000} Preece R.D. et al., 2000, ApJS, 126, 19 
\bibitem{Tavani1998} Tavani M., 1998, ApJ, 497, L21
\bibitem{Vietri1998} Vietri M. \& Stella L., 1998, ApJ, 507, L45 
\bibitem{Woosley1993} Woosley S. E., 1993, ApJ, 405, 273
\bibitem{Yi-ping Qin2001}  Yi-ping Qin et al., 2001, A\&A, 369, 537
\bibitem{Yonetoku2003} Yonetoku D. et al., astro-ph/0309217
\end{thebibliography}

\end{document}